\documentclass{article}
\usepackage{spconf,amsmath,graphicx,hyperref,amsfonts,amssymb}
\usepackage{booktabs}
\usepackage{algorithm}
\usepackage{algorithmic}

\def\y{{\mathbf y}}
\def\h{{\mathbf h}}
\def\a{{\mathbf a}}
\def\z{{\mathbf z}}
\def\W{{\mathbf W}}
\def\b{{\mathbf b}}
\def\E{{\mathcal E}}
\def\L{{\mathcal L}}
\def\R{{\mathbb R}}

\title{Decoder-only Conformer with Modality-aware\\
Sparse Mixtures of Experts for ASR}

\name{Jaeyoung Lee, Masato Mimura}
\address{NTT, Inc., Japan}

\begin{document}

\maketitle
\ninept

\begin{abstract}

We present a decoder-only Conformer for automatic speech recognition (ASR) that processes speech and text in a single stack without external speech encoders or pretrained large language models (LLM). The model uses a modality-aware sparse mixture of experts (MoE): disjoint expert pools for speech and text with hard routing and top-1 selection, embedded in hybrid-causality Conformer blocks (bidirectional for speech, causal for text). Training combines CTC on speech positions with label-smoothed cross-entropy for text generation. Our 113M-parameter model consistently improves WER over a 139M AED baseline on Librispeech (2.8\% vs.\ 3.2\% test-clean; 5.6\% vs.\ 6.0\% test-other). On Common Voice 16.1 with a single multilingual model across five languages, our approach reduces average WER from 12.2\% to 10.6\%. To our knowledge, this is the first randomly initialized decoder-only ASR that surpasses strong AED baselines via modality-aware routing and sparse MoE, achieving better accuracy with fewer active parameters and without alignment/adaptation modules.

\end{abstract}


\vspace{-0.1cm}
\section{Introduction}
\label{sec:intro}

Decoder-only architectures have recently gained attention in ASR \cite{gupta24_interspeech} due to their architectural similarities to large language models (LLMs). Decoder-only has the potential to unify speech-text modeling and to benefit from LLM pretraining, though it does not yet match the raw performance of attention encoder-decoder (AED) or transducer-based ASR systems. As for architectural backbones, Conformer \cite{gulati20_interspeech} remains the dominant choice for automatic speech recognition (ASR), typically deployed within encoder-decoder or transducer frameworks that achieve state-of-the-art performance.

A central challenge for decoder-only ASR is how to integrate heterogeneous modalities. 
Prior studies have approached this by explicitly aligning speech and text representations through modality adapters or length adapters \cite{fan2024alignformer,wang2023speech2textadapter,wu23_asru}. While effective, these strategies introduce additional modules and training complexity.

Sparse mixture-of-experts (MoE) \cite{shazeer2017moe,fedus2022switch} has emerged as an effective approach for scaling model capacity without proportional computational cost. In ASR, MoE has been applied primarily to encoder architectures \cite{mimura25_interspeech,you2021speechmoe}, and existing implementations do not explicitly partition experts by modality when processing both speech and text.

In this paper, we propose a decoder-only Conformer with modality-aware sparse MoE, which avoids explicit alignment or compression by allocating capacity separately for speech and text. Combined with hybrid causal/non-causal masking, this design enables continuous speech features and discrete text tokens to be processed directly within a single decoder stack, simplifying integration while maintaining temporal consistency.

Our contributions are as follows: (1) a unified decoder-only Conformer that handles both acoustic features and text tokens, (2) modality-aware expert pools with hard routing to better specialize capacity, and (3) a hybrid causality scheme that balances bidirectional speech modeling with causal text generation. Unlike recent decoder-only ASR work that depends on large pretrained speech encoders \cite{wang2023slm} or pretrained LLMs \cite{jia2024efficient_streaming_llm}, we show that a randomly initialized model can outperform encoder-decoder baselines. This is enabled by modality-aware MoE as the central mechanism, and we further improve training stability through auxiliary losses with CTC on speech positions and label-smoothed cross-entropy for text generation. To our knowledge, this combination has not been previously explored in decoder-only ASR \cite{gupta24_interspeech}.

\vspace{-0.1cm}
\section{Preliminaries}
\label{sec:related}

\begin{figure*}[ht!]
  \centering
  \includegraphics[width=\textwidth]{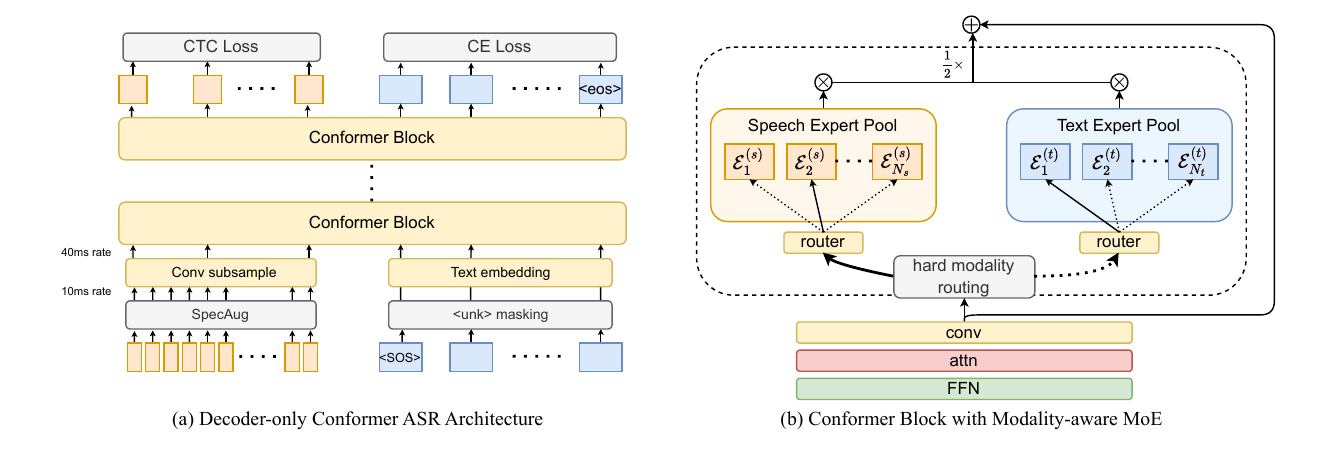}
  \vspace{-0.85cm}
  \caption{Overview of the proposed decoder-only Conformer with modality-aware MoE. Expert pools are partitioned by modality, where speech tokens route only to speech experts and text tokens to text experts. Each router chooses a single expert (top-1) within the corresponding pool.}
  \label{fig:archi}
\vspace{-0.2cm}
\end{figure*}

\vspace{-0.1cm}
\subsection{Decoder-only Architectures}
The decoder-only paradigm, which processes inputs through a single autoregressive model without separate encoder-decoder components, has emerged as a prevalent architecture in language modeling \cite{brown2020language,touvron2023llama}.

This architectural approach has subsequently been explored in multimodal speech-text contexts. Various methods have been proposed for incorporating speech signals: SpeechGPT \cite{zhang2023speechgpt} employs discretization of speech into tokens, TWIST \cite{hassid2024textually} explores speech-text manifold learning, and AudioPaLM \cite{rubenstein2023audiopalm} investigates fusion of separately trained speech and text models. Additional work in multimodal language models \cite{chu2023qwen,liu2024visual} has examined decoder-only architectures, predominantly utilizing discrete speech representations.

Within the specific domain of ASR, Gupta et al.\ \cite{gupta24_interspeech} investigated decoder-only Transformers for speech recognition, demonstrating that such architectures can process speech features and text tokens sequentially without explicit acoustic encoders. While speech discretization has been widely adopted in multimodal modeling, the necessity of this quantization step for ASR tasks remains an open question. In this work, we examine an alternative approach that maintains continuous speech representations alongside discrete text tokens, thereby preserving the original acoustic information without quantization-induced losses.

\vspace{-0.2cm}
\subsection{Conformer for Speech Recognition}

Conformer \cite{gulati20_interspeech} augments Transformer self-attention with depthwise separable convolutions, effectively capturing both global and local dependencies crucial for acoustic modeling. Various extensions have been proposed, including streaming variants \cite{chen2021developing,mimura2025_advancing}, efficient implementations \cite{burchi2021efficient}, and scaled versions \cite{zhang2022bigssl}. However, Conformer has primarily been deployed within encoder-decoder or transducer frameworks. Our work extends Conformer to decoder-only architectures with appropriate causal modifications.

\vspace{-0.2cm}
\subsection{Mixture-of-Experts in ASR and Multimodal Models}

Sparse MoE enables efficient model scaling through conditional computation \cite{shazeer2017moe}. The Switch Transformer \cite{fedus2022switch} simplified this approach with top-1 routing and demonstrated trillion-parameter scale training. In ASR, You et al.\ \cite{you2021speechmoe} introduced SpeechMoE for multi-domain adaptation, while Mimura et al.\ \cite{mimura25_interspeech} proposed Switch Conformer with universal phonetic experts for multilingual ASR.

In multimodal contexts, modality-aware expert partitioning has shown promise. MoMa \cite{lin2024moma} employs modality-specific expert groups in early fusion pretraining, while LIMoE \cite{mustafa2022limoe} demonstrates improved efficiency through language-image expert separation. VLMo \cite{bao2022vlmo} introduces modality experts with shared self-attention. Our approach uniquely combines these concepts within a decoder-only Conformer framework.

\vspace{-0.2cm}
\section{Methodology}
\label{sec:model}

\vspace{-0.1cm}
\subsection{Problem Formulation}

We consider acoustic features $\a = \{\a_1, \ldots, \a_{T_0}\} \in \R^{T_0 \times D_a}$, where $D_a$ is the feature dimension. These features are subsampled to length $T$ and paired with a target text sequence $\y = \{y_1, \ldots, y_U\}$ drawn from vocabulary $\mathcal{V}$. The task is to model the conditional probability $P(\y|\a)$ using a single decoder stack. Unlike encoder-decoder architectures that handle speech and text inputs separately, we form a unified input sequence.

\vspace{-0.2cm}
\subsection{Decoder-only Conformer Architecture}

\subsubsection{Speech Subsampling}
\label{sec:subsample}
Prior to projection to the main Conformer stack, acoustic features undergo $4\times$ temporal reduction through two stacked convolutional layers:
\begin{align}
\a^{(1)} &= \text{ReLU}(\text{Conv2D}(\a, \text{stride}=2)), \\
\a^{(2)} &= \text{ReLU}(\text{Conv2D}(\a^{(1)}, \text{stride}=2)),
\end{align}
producing a sequence of length $T = T_0/4$.

We do not use a separate audio encoder or CTC-based compressor \cite{wu23_asru}; this simple $4\times$ subsampling followed by a linear projection is the only front-end prior to the decoder stack.

\subsubsection{Input Representation}

We concatenate projected acoustic features and embedded text tokens into a single sequence:
\begin{equation}
\z = [\z_1^{(s)}, \ldots, \z_T^{(s)}, \z_1^{(t)}, \ldots, \z_{U-1}^{(t)}]
\end{equation}
where $\z_i^{(s)} = \W_{\text{proj}}\a_i + \b_{\text{proj}}$ for speech positions and $\z_j^{(t)} = \text{Embed}(y_j)$ for text positions. Positional encodings are added to each element:
\begin{equation}
\tilde{\z}_i = \begin{cases}
\z_i^{(s)} + \text{PE}(i) & \text{if } i \leq T \\
\z_{i-T}^{(t)} + \text{PE}(i) & \text{if } i > T
\end{cases}
\end{equation}
where $\text{PE}(\cdot)$ denotes sinusoidal positional encoding. Since speech features occupy positions $[1,T]$ and text tokens begin from position $T+1$, the boundary between modalities is known deterministically.

\vspace{-0.2cm}
\subsection{Modality-aware Sparse Mixture-of-Experts}
\label{sec:moe}

Integrating heterogeneous modalities within a single decoder-only stack poses two challenges: the mismatch between continuous speech features and discrete text embeddings and large differences in sequence length. Prior work addresses these with modality-matching adapters or length-reduction modules, adding components and training complexity. We instead embed a modality-aware sparse MoE within Conformer blocks to partition capacity by modality, bypassing the need for explicit alignment or length adaptation while preserving acoustic detail.

\vspace{-0.3cm}
\subsubsection{Hybrid Causal Conformer Blocks}
\label{sec:causal_mask}

Each of the $L$ Conformer blocks applies modality-dependent operations, combining bidirectional modeling for speech with causal generation for text. For position $i$ at layer $l$:

\vspace{-0.1cm}
\begin{align}
\h_i^{(l)} = \text{ConformerBlock}(\h_i^{(l-1)}, \mathcal{M}_i),
\end{align}

\vspace{-0.1cm}
where $\mathcal{M}_i$ determines the attention mask and convolution type based on modality. The attention mask follows \cite{gupta24_interspeech}:

\begin{equation}
M_{ij} = \begin{cases}
1 & i,j \leq T \quad \text{(speech attending to all speech)} \\
1 & i > T, j \leq T \quad \text{(text attending to all speech)} \\
1 & i > T, j > T, j \leq i \quad \text{(causal text)} \\
0 & \text{otherwise.}
\end{cases}
\end{equation}

\begin{figure}[ht!]
  \centering
  \includegraphics[width=0.95\columnwidth]{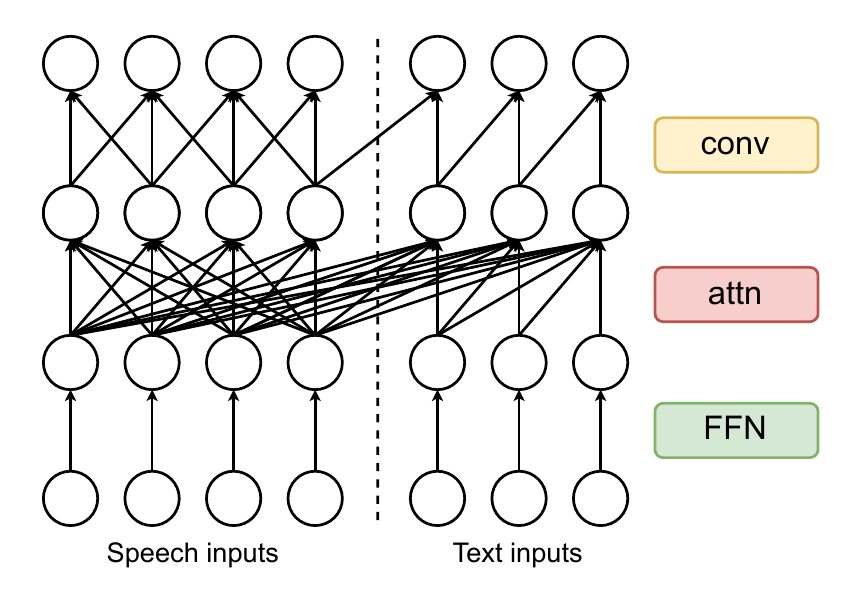}
  \vspace{-0.7cm}
  \caption{Non-causal and causal masks are applied to speech and text representations, respectively, across all convolution and self-attention layers.}
  \label{fig:causal}
  \vspace{-0.3cm}
\end{figure}

Thus, speech representations attend bidirectionally to all other speech frames ($1\!:\!T$), while each text token attends to the full speech sequence plus all preceding text tokens up to position $i$ (Figure~\ref{fig:causal}).

Each block applies convolution with kernel size 15 and modality-dependent receptive fields: speech positions use the full 15-frame window, while text positions are limited to the most recent 8 frames in a causal manner. Convolutional filters are shared across modalities, and batch normalization is replaced by layer normalization for causality in text.

The overall block structure follows the standard Conformer \cite{gulati20_interspeech}, except that the second feed-forward layer is replaced with our modality-aware MoE:
\begin{align}
\h_i^{[1]} &= \h_i + \tfrac{1}{2}\text{FFN}(\h_i), \\
\h_i^{[2]} &= \h_i^{[1]} + \text{MHSA}(\h_i^{[1]}, \mathcal{M}_i), \\
\h_i^{[3]} &= \h_i^{[2]} + \text{Conv}(\h_i^{[2]}, \mathcal{M}_i), \\
\h_i^{[4]} &= \h_i^{[3]} + \tfrac{1}{2}\text{MoE}(\h_i^{[3]}, \mathcal{M}_i),
\end{align}
where $\h_i^{[4]}$ is the block output.

\begin{table*}[ht!]
\centering
\caption{LibriSpeech WER (\%). The AED baseline has more active parameters (139M) than the proposed decoder-only model (113M).}
\vspace{0.1cm}
\label{tab:librispeech_detailed}
\begin{tabular}{lcccc}
\toprule
\textbf{Model} & \textbf{Backbone} & \textbf{\# Active Params} & \textbf{test-clean} & \textbf{test-other} \\
\midrule
AED, 17-layer enc / 6-layer dec & Conformer & 139M & 3.2 & 6.0 \\
dec-only 17-layer & Transformer & 64M & 3.6 & 7.8 \\
dec-only 17-layer & Conformer & 113M & 3.4 & 6.4 \\
w/ MoE (no modality grouping, top-2) & Conformer & 113M & \textbf{2.8} & 6.3 \\
\textbf{w/ MoE, modality-aware (top-1 per modality)} & \textbf{Conformer} & \textbf{113M} & \textbf{2.8} & \textbf{5.6} \\
\bottomrule
\end{tabular}
\end{table*}

\begin{table*}[ht!]
\centering
\vspace{-0.3cm}
\caption{Common Voice 16.1 (de, en, es, fr, it) WER (\%) and average.}
\vspace{0.1cm}
\label{tab:cv_results}
\begin{tabular}{lcccccccc}
\toprule
\textbf{Model} & \textbf{Backbone} & \textbf{\# Active Params} & \textbf{de} & \textbf{en} & \textbf{es} & \textbf{fr} & \textbf{it} & \textbf{avg} \\
\midrule
AED, 17-layer enc / 6-layer dec & Conformer & 139M & 9.3 & 17.8 & 9.2 & 14.1 & 10.5 & 12.2 \\
dec-only 17-layer & Transformer & 64M & 12.5 & 21.9 & 12.0 & 17.5 & 14.4 & 15.7 \\
dec-only 17-layer & Conformer & 113M & 10.1 & 18.9 & 10.0 & 15.0 & 11.8 & 13.2 \\
w/ MoE (no modality grouping, top-2) & Conformer & 113M & 8.4 & 16.6 & 8.3 & 13.1 & 9.8 & 11.2 \\
\textbf{w/ MoE, modality-aware (top-1 per modality)} & \textbf{Conformer} & \textbf{113M} & \textbf{7.8} & \textbf{16.0} & \textbf{7.8} & \textbf{12.3} & \textbf{9.1} & \textbf{10.6} \\
\bottomrule
\vspace{-0.7cm}
\end{tabular}
\end{table*}

\vspace{-0.2cm}
\subsubsection{Expert Pool Partitioning}

We partition experts into two modality-specific expert pools:
\begin{equation}
\E = \E^{(s)} \cup \E^{(t)}, \quad \E^{(s)} \cap \E^{(t)} = \emptyset
\end{equation}
where $\E^{(s)} = \{E_1^{(s)}, \ldots, E_{N_s}^{(s)}\}$ contains $N_s$ speech experts and $\E^{(t)} = \{E_1^{(t)}, \ldots, E_{N_t}^{(t)}\}$ contains $N_t$ text experts.

\subsubsection{Routing Mechanism}

For input $\h_i$ at position $i$, the active expert pool is determined by modality:
\begin{equation}
\E_{\text{active}}(i) = \begin{cases}
\E^{(s)} & \text{if } i \leq T, \\
\E^{(t)} & \text{if } i > T.
\end{cases}
\end{equation}

Within the active pool, a learned router $g(\cdot)$ computes selection probabilities:
\begin{equation}
\mathbf{p}_i = \text{Softmax}(\W_g^{(m)}\h_i + \b_g^{(m)})
\end{equation}
where $m \in \{s, t\}$ indicates the modality, and $\W_g^{(m)} \in \R^{N_m \times D_{\text{model}}}$.

Expert selection follows the top-$k$ strategy:
\begin{equation}
\mathcal{S}_i = \text{TopK}(\mathbf{p}_i, k)
\end{equation}

with the output aggregated as
\begin{equation}
\text{MoE}(\h_i) = \sum_{j \in \mathcal{S}_i} p_{i,j} \cdot E_j(\h_i)
\end{equation}

We use $k=2$ for the single-pool MoE baseline and $k=1$ within each modality-specific pool for our method.

\subsubsection{Load Balancing}

Sparse MoE layers can suffer from \emph{routing collapse}, where most tokens are 
sent to a few experts while others are rarely used. To encourage more uniform 
utilization, we add an auxiliary load-balancing loss \cite{fedus2022switch}:

\begin{equation}
\L_{\text{balance}} = \alpha \sum_{m \in \{s,t\}} \sum_{j=1}^{N_m} f_j^{(m)} \, P_j^{(m)}
\end{equation}

where $f_j^{(m)}$ is the fraction of tokens routed to expert $j$ (empirical load) 
and $P_j^{(m)}$ is the mean router probability for that expert. This loss nudges 
the router toward distributing tokens more evenly across experts. $\alpha$ is a 
hyperparameter.

\subsection{Training Objective}

The model employs a multi-task learning approach with three loss components. First, we apply CTC loss exclusively to speech positions, a technique not previously explored in decoder-only ASR that significantly improves performance:
\begin{equation}
\L_{\text{CTC}} = -\log P_{\text{CTC}}(\y|\h_{1:T}^{(L)})
\end{equation}
where $\h_{1:T}^{(L)}$ are the final-layer representations for speech positions.

Second, cross-entropy loss with label smoothing ($\epsilon=0.1$) is computed for text positions:
\begin{equation}
\L_{\text{CE}} = -\sum_{i=1}^{U} (1-\epsilon)\log P(y_{i+1}|\z_{\leq T+i}) + \epsilon \log \frac{1}{|\mathcal{V}|}
\end{equation}

The total training objective combines these with the load-balancing loss:
\begin{equation}
\L = \L_{\text{CE}} + \alpha_{\text{CTC}} \L_{\text{CTC}} + \alpha \L_{\text{balance}}
\end{equation}
where $\alpha_{\text{CTC}}=0.3$ and $\alpha=0.1$. The addition of CTC loss to speech positions and label smoothing to text generation, absent in prior decoder-only work \cite{gupta24_interspeech}, proves crucial for achieving competitive performance.

\vspace{-0.2cm}
\subsection{Inference}

During inference, we employ an offline decoding strategy where all acoustic features are first processed through the model to generate speech representations. These representations are then cached, and the model autoregressively generates text tokens conditioned on the complete speech context. We use beam search with beam size 4 for all experiments.

\section{Experimental Evaluations}
\label{sec:exp}

\vspace{-0.2cm}
\subsection{Datasets and Preprocessing}
We evaluate our models on both LibriSpeech and Common Voice 16.1. For LibriSpeech \cite{panayotov2015librispeech}, we use the standard 960-hour training set, with dev-clean and dev-other subsets for validation, and test-clean and test-other subsets for evaluation. For Common Voice 16.1 \cite{ardila2020commonvoice}, we construct a multilingual subset covering five languages - German (de), English (en), Spanish (es), French (fr), and Italian (it) - using the validated splits to ensure data quality. For all experiments, we extract 80-dimensional log-Mel spectrogram features computed with a 25 ms window and a 10 ms hop length. The transcriptions are tokenized using byte-pair encoding (BPE) with a vocabulary size of 2k.

\vspace{-0.1cm}
\subsection{Model Configurations}

Our proposed model is a 17-layer decoder-only Conformer with a hidden size of $D_{\text{model}}=512$, eight attention heads, and a base feed-forward network (FFN) dimension of 2048. We investigate two MoE configurations to replace the second FFN in each Conformer block. The first configuration employs a standard MoE approach with 16 experts with top-2 routing, selecting two experts per token regardless of modality. The second configuration implements our modality-aware MoE with two separate expert pools: 8 experts for speech and 8 for text, using top-1 routing within the appropriate pool based on token position. Both configurations employ hybrid causal masks as described in \ref{sec:causal_mask}. To keep the \emph{active} compute comparable across configurations, each MoE expert uses an FFN dimension of 1024 (half the original 2048), so the active parameter budget remains 113M, while the \emph{total} number of parameters in the MoE variants is 220M (only a subset of experts is active per token).

For comparison, we evaluate against several baselines. The first is a sequence-to-sequence Conformer with a 17-layer encoder and 6-layer decoder (139M active parameters). The second is a decoder-only Transformer with 17 layers and 64M parameters, following the setup in \cite{gupta24_interspeech}. We also include a decoder-only Conformer with 17 layers but without MoE (113M parameters). All models share identical training objectives: CTC auxiliary loss (weight 0.3) on speech representations and cross-entropy loss with label smoothing (weight 0.1) for text generation, ensuring fair comparison across architectures.

\vspace{-0.1cm}
\subsection{Training Details}

Models are trained using the Adam optimizer \cite{kingma2014adam} with $\beta_1=0.9$, $\beta_2=0.999$, and peak learning rate of $1.5 \times 10^{-3}$. We employ the same learning rate schedule as \cite{vaswani2017attention}, with 25,000 warmup steps followed by inverse square root decay. Batch size is set to 50 minutes. For data augmentation, SpecAugment \cite{park2019specaugment} is applied to speech features, and word dropout is applied to text tokens, masking each token with \texttt{<unk>} with probability 0.125.

Training epochs vary by dataset: Common Voice models train for 30 epochs with the final 5 epochs averaged for evaluation, while LibriSpeech models train for 50 epochs with the final 10 averaged. For LibriSpeech only, we apply speed perturbation \cite{ko2015audio} with factors of 0.9, 1.0, and 1.1 to augment the 960-hour training data.

All Conformer models employ a 4× subsampling rate for speech through convolutional layers.

\subsection{Results}
\label{sec:results}

Table~\ref{tab:librispeech_detailed} shows that the modality-aware decoder-only Conformer achieves 2.8\% / 5.6\% WER on \textit{test-clean/other}, improving over the larger AED baseline (3.2\% / 6.0\%). Within decoder-only models, MoE also brings clear gains over the non-MoE backbone (3.4\% / 6.4\%), with modality partitioning further improving robustness on \textit{test-other} compared to a single-pool MoE.

Table~\ref{tab:cv_results} reports consistent improvements across five Common Voice languages, reducing average WER from 12.2\% to 10.6\%. Taken together with the LibriSpeech results, these findings indicate that the introduction of sparse MoE contributes substantially to the observed gains. 
While prior ASR work has mainly leveraged MoE for \emph{multilingual} capacity 
expansion, our formulation extends it to a \emph{multi-modal} setting, where 
speech and text experts are explicitly separated. The results verify that such 
modality-aware partitioning works effectively, providing a generally applicable 
mechanism for scaling decoder-only ASR without increasing active parameters.

\vspace{-0.1cm}
\section{Conclusion}
\label{sec:concl}

This paper introduced a decoder-only Conformer architecture with modality-aware sparse mixture-of-experts for automatic speech recognition. By partitioning experts by modality and implementing hybrid causality, the model effectively balances acoustic and linguistic modeling within a unified framework. Experimental results demonstrate substantial improvements over strong baselines while maintaining parameter efficiency.

Future work includes extending this framework to streaming ASR through 
chunk-wise processing, and investigating learned routing mechanisms that allow 
flexible sharing or transfer between modality-specific expert pools.

\clearpage

\bibliographystyle{IEEEbib}
\bibliography{strings}

\end{document}